\documentstyle[12pt,twoside,fleqn,espcrc1,epsfig]{article}

\def\bge{\begin{equation}}
\def\ene{\end{equation}}
\def\bg{\begin{eqnarray}}
\def\en{\end{eqnarray}}

\newcommand{\AmS}{{\protect\the\textfont2
  A\kern-.1667em\lower.5ex\hbox{M}\kern-.125emS}}

\title{Hadrons in Dense Matter} 

\author{A. W. Thomas\address{Department of Physics and Mathematical
Physics, \\
and Special Research Centre for the Subatomic Structure of Matter, \\
        University of Adelaide,  \\ 
        Adelaide, Australia 5005}}
\begin{document}
\maketitle
\vspace{-8cm}
\begin{flushright}
{\footnotesize 
Invited talk at the Int. Conf. on Quark Lepton Nuclear Physics \\
Osaka, May 20 - 23, 1997 \\
ADP-97-20/T257}
\end{flushright}
\vspace{7cm}
\begin{abstract}
There is currently enormous interest in the investigation of how hadron
properties may be altered by immersion in matter. There is strong
evidence of a reduction in the mass of the rho meson from relativistic
heavy ion collisions as well as a hint from a recent experiment on
photoproduction in light nuclei.
We briefly review the main theoretical ideas which lead one to
expect the mass of a hadron to change in matter, including the
various QCD-based methods, notably the QCD sum rules, as well as
mean-field, quark based models like QMC and conventional nuclear
approaches such as QHD.
\end{abstract}

\section{INTRODUCTION}

Over the past decade there has been a great deal of theoretical work
directed at understanding how hadronic properties change in nuclear
matter. Recently this activity has been stimulated by the evidence from
relativistic heavy ion collisions that the mass of the $\rho$ meson in
matter may be several hundred MeV lower at 2-3 times
nuclear matter density. We begin with a brief reminder of this
experimental evidence. Then we turn to recent theoretical developments,
including QHD, the quark meson coupling model (QMC) and QCD sum rules,
including the effect of absorptive channels. In conclusion we mention
recent experimental work which aims to detect the shift of the $\rho$
mass in light nuclei using sub-threshold photo-production.

\section{RELATIVISTIC HEAVY ION COLLISIONS}

There is a variety of experiments underway at CERN (CERES and HELIOS-3)
and GSI (HADES) and planned at
RHIC which aim to produce high density hadronic matter through the
collision of relativistic heavy ions. In the collisions of $S$ on $Au$
at SPS/CERN at 200 GeV \cite{ceres} there is a large excess of $e^+e^-$
pairs in the invariant mass region around 400 MeV. Numerical simulations
of the collision dynamics suggest that the region of hot dense matter
formed in such a collision (the ``fire cylinder'') should occupy a
volume about 7fm in diameter and 4fm long. At formation it should have a
peak density around $2-3\rho_0$ (with $\rho_0$ the saturation density
of symmetric nuclear matter) and contain roughly 100 baryons, 100 pions
and perhaps 50 $\rho$ and $\omega$ mesons \cite{LKB}. The most
straightforward explanation of the excess lepton pairs is that the mass
of the $\rho$ meson in matter at such densities is lowered to between
400 and 600 MeV \cite{L2}.

Of course, the dynamics of matter under such extreme conditions is very
much an unknown area and there are alternative mechanisms which have
been suggested to explain the excess. For example, in relation to the
CERES data, Chanfray {\it et al.} \cite{CRW}
have emphasised the importance of
channels such as $\pi N$ and $\pi \Delta$ which may produce an
enhancement in the number of pions in the system. These can then
annihilate into lepton pairs. It may be some time before there is a
consensus on the mechanism involved and new data, particularly involving
systems such as $Pb$ on $Pb$, which should produce an even larger fire
cylinder, will play a vital role. Nevertheless, the theoretical efforts
to understand the behaviour of hadrons in dense matter have received a
tremendous stimulus because of the enhancement in precisely the region
one expects if the $\rho$ mass decreases in the way many theoretical
approaches have suggested.

\section{THEORETICAL DEVELOPMENTS}

A great deal of the recent theoretical on the possible decrease of
hadron masses in nuclear matter has also been stimulated by the
suggestion of Brown-Rho scaling \cite{BR}. Starting with a chiral
soliton model of nuclear matter these authors suggested the following
behaviour for the variation of key parameters with density:
\begin{equation}
\frac{m_V^*}{m_V} \approx \frac{m_N^*}{m_N} \approx \frac{m_\sigma^*}
{m_\sigma} \approx \left( \frac{\langle r^2 \rangle_N}{\langle r^2 \rangle_N^*}
\right)^{\frac{1}{2}} \approx \frac{f^*_\pi}{f_\pi}.
\label{br}
\end{equation}
As $\frac{f^*_\pi}{f_\pi}$ is expected to vanish as chiral symmetry is
restored at high density all of the masses in Eq.(\ref{br}) (i.e., the
vector meson masses, $m_V$, the nucleon mass $m_N$ and the $\sigma$
meson mass $m_\sigma$) are expected to decrease as the density
increases. Taking as a guide the reduction of the ratio $m_N^*/m_N$
calculated in QHD, that is around 0.6 at $\rho_0$, one expects the
$\rho$ mass to be as low as 500 MeV, even at normal nuclear matter
density.

On the other hand, there are very strict limits on the possible
variation of the size of the nucleon in nuclear matter \cite{sick}. 
The mean square radius of the nucleon cannot vary by more than perhaps
5\% and Eq.(\ref{br}) is in clear contradiction with such a limit. 
It remains to be seen whether it is possible to generalise the Brown-Rho
analysis to relax the condition on the nucleon radius while retaining
the results for the masses. In any case, the main point of that work is
the stimulus it has provided to the field.

\subsection{Quantum Hadrodynamics}

Quantum hadrodynamics (QHD) developed out of the wish to have a
renormalizable, covariant theoretical framework for nuclear physics
\cite{wal}. In its simplest form, QHD involves a system of point-like,
Dirac nucleons coupled to elementary scalar ($\sigma$) and vector
($\omega$) isoscalar mesons. In Hartree approximation the self
consistent solution for nuclear matter leads to a nucleon effective mass
at saturation density of about 60\% of the free mass. Going beyond
Hartree approximation tends to yield a value more in the range 
70--80\%.

The first study of the mass of a vector meson (in this case the
$\omega$) within QHD was by Saito, Maruyama and Soutome \cite{sms}. For
a recent review we refer to the work of Cohen {\it et al.}
\cite{cohen}. All of the studies within QHD tend to lead to a lowering
of the mass of the $\omega$, mainly because of the coupling to
$N\bar{N}$ states. (Without that the effect of particle-hole excitations
would be to raise the mass.) This has led Saito and Thomas to
question the sign of the effect in a QHD-type model in the case where
the coupling to $N\bar{N}$ states is suppressed by the finite size of
the nucleon \cite{st1}.

One other aspect of the QHD treatment that has so far not been
sufficiently appreciated is the finding of Jean, Piekarewicz and
Williams \cite{jpw} that the shift of the $\omega$ mass is not only
momentum dependent but that the shift is different for transversely
polarized and longitudinally polarized mesons 
(i.e., $m_V^{*T} \neq m_V^{*L}$). (This was also emphasised more
recently by Eletsky and Ioffe \cite{EI}.) The latter could be particularly
important for the analysis of real experimental data. For a vector meson
at rest these authors found a very similar mass shift to that in earlier
work, with the effective mass of the meson about 80\% of its free value
at nuclear matter density.

\subsection{The Quark Meson Coupling Model}

A truly consistent theory capable of describing the transition from
meson and baryon degrees of freedom to quarks and gluons might be
expected to incorporate the internal quark and gluon degrees of freedom
of the particles themselves. The quark meson coupling (QMC) model was
originally suggested by Guichon \cite{guichon} in order to investigate
precisely these effects. It has since been developed extensively
\cite{hadron,yazaki,jenn,oth} for nuclear matter and extended to
provide a very realistic description of finite nuclei
\cite{finite1,finite2,bm}.

Within the QMC model the properties of nuclear matter 
are determined by the self-consistent coupling of scalar ($\sigma$) and
vector ($\omega$) fields to the {\it quarks within the nucleons},
rather than to the nucleons themselves.
As a result of the scalar coupling the internal structure of the nucleon
is modified with respect to the free case. In particular, the small mass
of the quark means that the lower component of its wave function
responds rapidly to the $\sigma$ field, with a consequent decrease in
the scalar density. As the scalar density is itself the source of the
$\sigma$ field this provides a mechanism for the saturation of nuclear
matter where the quark structure plays a vital role.

In a simple version of the model, 
where nuclear matter was considered as a collection of
static, non-overlapping bags it was shown that a satisfactory
description of the bulk properties of nuclear matter can be
obtained~\cite{st1,guichon}.
Of particular interest is the fact that the extra degrees of freedom,
corresponding to the internal structure of the nucleon, result in a
lower value of the incompressibility of nuclear matter than obtained in
approaches based on point-like nucleons -- such as
QHD~\cite{wal} -- at least at the same level of sophistication
(Hartree approximation).  In fact, the prediction is in
agreement with the experimental value once the binding energy and
saturation density are fixed. Improvements to the model, including the
addition of Fermi motion, have not
altered the dominant saturation mechanism.  Furthermore, it is possible
to give a clear understanding of the relationship between this model and
QHD~\cite{ST}. Surprisingly the model seems 
to provide a semi-quantitative explanation of the Okamoto-Nolen-Schiffer
anomaly when quark mass differences are included~\cite{ok}.
Finally the model has been
applied to the case where quark degrees of freedom
are undisputedly involved -- namely the nuclear EMC effect~\cite{st2}.

Our main interest is in the predictions of the model for the variation
of hadron masses in dense matter. As the $\omega$ and $\rho$ are simple
$q\bar{q}$ states in QCD, the QMC model has been generalised
to allow for the self-consistent determination of the vector meson
masses as well as the nucleon mass \cite{var}. The variation of the
resulting nucleon and $\omega$ masses are shown in Figs. 1 and 2 
as a function of the density of symmetric nuclear matter
\cite{var}.
\begin{figure}[hbt]
\begin{center}
\epsfig{file=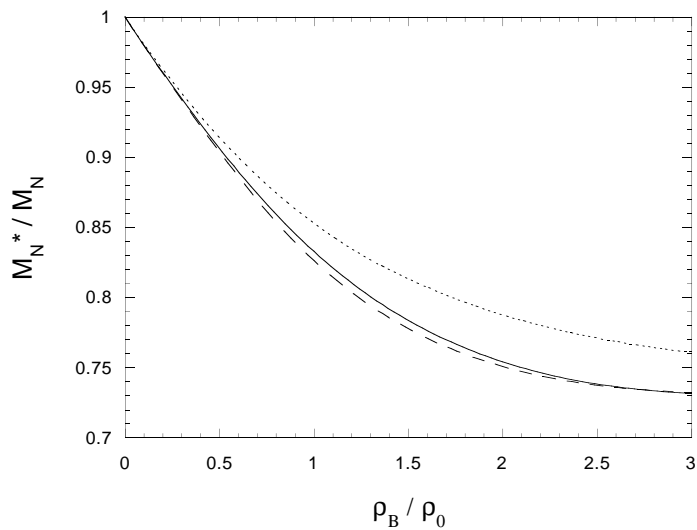,height=7cm}
\caption{Effective mass of the nucleon in symmetric nuclear matter.
The three curves correspond to three parametrizations of the $\sigma$
mass as a function of density -- see Ref.[24].}
\label{fig:enm}
\end{center}
\end{figure}
At low density the behaviour of these mass variations is well
approximated by a term linear in the density:
\begin{equation}
\left( \frac{M_N^{\star}}{M_N} \right) \simeq 1 - 0.21
\left( \frac{\rho_B}{\rho_0} \right), 
\label{nstr2}
\end{equation}
and
\bge
\left( \frac{m_v^{\star}}{m_v} \right) \simeq 1 -
0.17 \left( \frac{\rho_B}{\rho_0} \right).
\label{vmm2}
\ene

Using the QMC model it was also possible to study the variation of the
masses of the $\Lambda, \Sigma$ and $\Xi$ as well. From these
calculations we were led to a new, simple scaling relation between the
hadron masses \cite{var}:
\bge
\left( \frac{\delta m_v^{\star}}{\delta M_N^{\star}} \right) \simeq
\left( \frac{\delta M_\Lambda^{\star}}{\delta M_N^{\star}} \right)
\simeq
\left( \frac{\delta M_\Sigma^{\star}}{\delta M_N^{\star}} \right) \simeq
\frac{2}{3} \ \ \ \mbox{ and } \ \ \
\left( \frac{\delta M_\Xi^{\star}}{\delta M_N^{\star}} \right) \simeq
\frac{1}{3} ,
\label{scale}
\ene
where $\delta M_j^{\star} \equiv M_j - M_j^{\star}$.  The factors,
$\frac{2}{3}$
and $\frac{1}{3}$, in Eq.(\ref{scale}) come from the ratio of the number
of non-strange quarks in $j$ to that in the nucleon.  This means that the
hadron mass is primarily
determined by the number of non-strange quarks. These experience 
the common scalar field generated by surrounding nucleons in
medium, and the strength of the scalar field.

\begin{figure}[ht]
\begin{center}
\epsfig{file=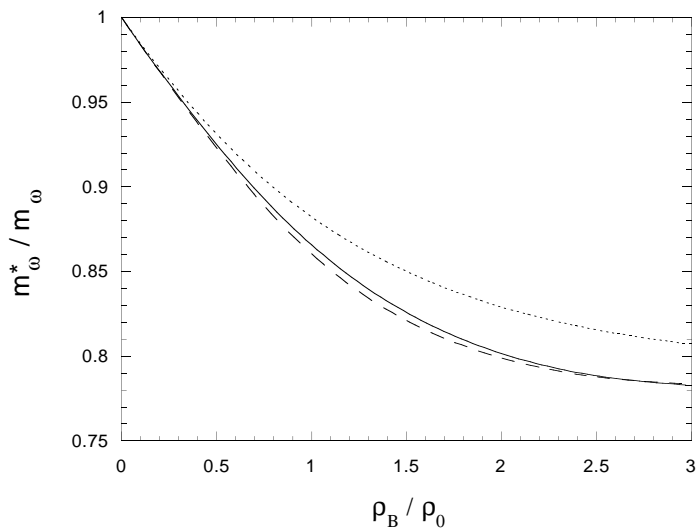,height=7cm}
\caption{Effective ($\rho$- or) $\omega$-meson mass in symmetric nuclear
matter -- from Ref.[24].}
\label{fig:eom}
\end{center}
\end{figure}

In view of the interest in the CERES results noted earlier, it is
interesting to estimate the shift in the mass of the $\rho$ meson
predicted by the QMC model at the appropriate densities. At $\rho_0$ we
find $m_\rho - m_\rho^* \approx 140$MeV, while at $2\rho_0$ we find 
$m_\rho^* \approx 580$MeV. The latter is certainly in the right
range needed to understand the excess leptons observed in the
experiment.

\subsubsection{Constraints on the QMC model}

We have already noted the strong constraints on the possible variation
of the internal structure of the nucleon from $y$-scaling \cite{sick}. 
It has only
recently proven possible to accurately calculate the variation of the
nucleon electromagnetic form factors within the QMC model and check that
they are consistent with existing constraints \cite{LU}. Figure 3 shows the
results and we see that the variation of the magnetic form factors
(which tend to be the most important) is particularly small.
\begin{figure}[htb]
\centering{\
\epsfig{file=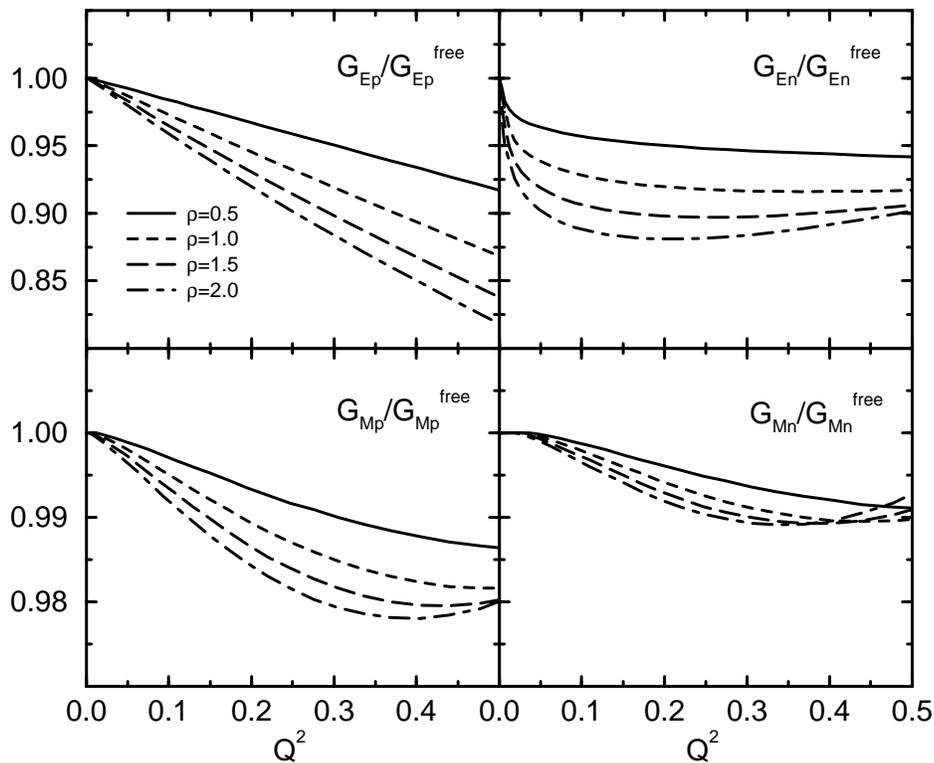,height=12cm}
\caption{The nucleon electromagnetic form factors
in the nuclear medium (calculated within the QMC model) 
relative to those in free space case -- from
Ref.[25].}
\label{fig3.ps}}
\end{figure}

\subsubsection{Possible connection to QCD}

The saturation mechanism within the QMC model is rather more general
than the bag upon which it was based. Indeed, the details of the
internal structure of the nucleon do not appear in the final equations,
all one needs is the density dependence of the coupling constant of the
scalar meson to the ``nucleon''. It would be extremely interesting to
see whether this could be calculated directly from QCD -either on the
lattice or through the Dyson-Schwinger approach. If it were possible to
input this variation from such a calculation one would, in a very real
sense, have a means to calculate nuclear properties from QCD itself.

\section{QCD SUM RULES}

The QCD sum rule technique is used to relate two evaluations of a
current-current correlator. On one side, the correlator is evaluated
using the operator product approach to QCD to yield an expression which
is valid at large momentum transfer. For example, if we write the
$\rho-\rho$ correlator, $\Pi^{\mu \nu}$, as $(g^{\mu \nu} - q^\mu
q^\nu/q^2)\Pi$, then (using $-q^2 = Q^2 > 0$) one finds:
\begin{equation}
\frac{12\pi}{Q^2} \Pi(Q^2) = \frac{d}{\pi} \left[ -c_0
ln\frac{Q^2}{\mu^2} + \frac{c_1}{Q^2} + \frac{c_2}{Q^4} + \ldots
\right].
\label{lhs}
\end{equation}
In Eq.(\ref{lhs}) the term in $1/Q^2$ involves the quark masses, the
$1/Q^4$ term the quark and gluon condensates, the $1/Q^6$ term the
4-quark condensate, and so on. On the other side, one writes a
dispersion relation in which the $\rho$ meson is often approximated by a
simple pole and the continuum by a simple $\theta$-function. The Borel
transform is often employed to make the region over which the two sides
can be equated large enough to check whether they match well.

In free space the QCD sum rule technique appears to work well and the
shape of the imaginary part of $\Pi$ ($Im \Pi$) is relatively well
described by vector meson dominance. In matter one needs to allow for
the variation of the various condensates and to modify the dispersion
relation to describe the hadron in medium \cite{Druk}. Hatsuda {\it et al.}
concluded that the mass of the $\rho$ should behave in a very similar
manner to that found in QMC (c.f., Eq.(\ref{vmm2}) above) \cite{hl}:
\bge
\frac{m_V^*}{m_V} \approx 1 - 0.18 \frac{\rho}{\rho_0}.
\label{HAT}
\ene
Jin and Leinweber found an almost identical result \cite{lj}, but with somewhat
larger errors, corresponding to a realistic estimate of the current state
of knowledge of all the relevant parameters, $m_V^*/m_V = 0.78 \pm
0.08$. 

An important new observation concerning this problem was made recently
by Klingl {\it et al.} \cite{kkw} -- see also Asakawa and Ko \cite{AK}.
These authors estimated the effect of the coupling of the $\rho (\omega) N$
system to $\pi N, \pi \pi N, \Sigma K$ etc. Using a phenomenological
Lagrangian based on chiral $SU(3)$ symmetry they showed that the affect
of absorption could be very strong.
In the case of the $\rho$, the shape of the 
$Im \Pi(s)$ usually used was shown to be quite wrong. While the spectral
function was {\it much} broader and appeared to peak at a 
much lower invariant mass it was clear from the real part that the mass
of the $\rho$ had not shifted significantly.

The analysis of Klingl {\it et al.} omitted the mean-field scalar
attraction that usually leads to a lowering of the mass of a hadron in
matter. It also omitted the effect of s-channel, baryon resonances which could
be important in some circumstances. Nevertheless, this work has served
to remind people working in this field that the very notion of a hadron
``mass'' in medium is a model dependent concept. One is really looking
at the response of the medium in a specific channel.

\section{AN ALTERNATIVE EXPERIMENT}

While most attention is focussed on the experiments involving
relativistic heavy ions and matter under extreme conditions, it would
also be very important to measure a shift in normal nuclei. An attempt
to do just this is currently underway at TJNAF (following the initial
proposal by Bertin and Guichon \cite{BG,Grho}) and at INS \cite{ins}.
The INS experiment aimed to measure sub-threshold $\rho^0$
photoproduction on light nuclei such as $^{3,4}He$ and $^{12}C$ using
the TAGX spectrometer at the 1.3GeV Tokyo Electron Synchrotron. Light
nuclei offer the advantage of less final state interactions for the
outgoing $\pi^+\pi^-$ pairs. Initial results presented at this meeting
suggest evidence that the mass of the $\rho$ may be somewhat lower than
the free case, even for $^3He$ \cite{Mar}.

\begin{figure}[hbt]
\begin{center}
\epsfig{file=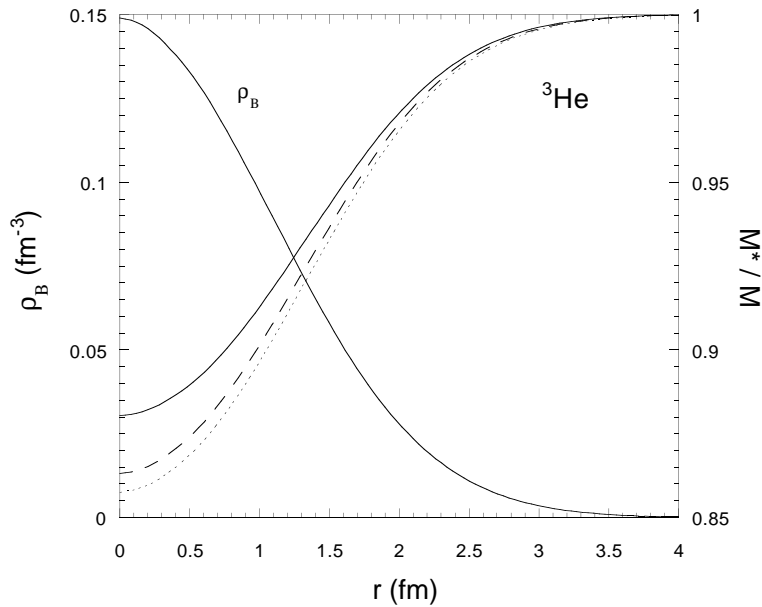,height=8cm}
\caption{Effective $\rho$-meson mass and the density distribution in
$^3$He. The solid, dashed and dotted curves correspond to the same
parameter sets used in Figs. 1 and 2 (above) -- from Ref.[36].}
\label{fig:3he}
\end{center}
\end{figure}
In order to estimate the size of the effects expected in the TAGX
experiment, Saito {\it et al.} \cite{light} 
have used the latest version of the QMC
model, in which the meson masses are self-consistently obtained in the
mean scalar field of the nucleus \cite{var}. As the mean field
description is not good for a nucleus as light as $^3He$ the effective
mass as a function of position in the nucleus was estimated in local
density approximation using a phenomenological density distribution.
Both the form used and the effective mass of the $\rho$ are shown in
Fig. 4. On average the mass of the $\rho$ is shifted down by about 40MeV
in this case. It will be very interesting to see the final results of
this experiment for both the coherent and incoherent $\rho$ production.
A comparison with data on a slightly heavier nucleus such as $^{12}C$
would also be very useful.

In the light of the results of Klingl {\it et al.} \cite{kkw}, 
discussed earlier, it will also be important to improve this first
estimate by estimating the effect of coupled, absorptive channels.
This is also clearly relevant to the TJNAF experiment, where the
comparison of the rather different behaviour expected of the $\rho$ and
$\omega$ was anticipated to provide an important signal \cite{BG}.

\section{CONCLUSION}

The study of the variation of hadron properties in matter is fundamental
to our developing understanding of the strong interaction. As
temperature and density vary we expect to see a shift in the relative
importance of quark and hadron degrees of freedom. There are already
tantalising hints from CERES that the mass of the $\rho$ meson may decrease
rather dramatically with increasing density. 
Even in a nucleus as light as $^3He$,
the TAGX experiment suggests that it may also
decrease. We may expect a great deal more experimental information in
the coming years.

The theoretical consensus seems to be that without channel coupling
(absorptive) effects the mass of a vector meson should decrease by about
150--200 MeV at $\rho_0$. The situation with respect to the mass of
virtual $\rho$ and $\omega$ mesons is much less clear. One might expect
the effect of channel coupling to be attractive in this case.
Nor is there a
consensus on the behaviour of the $\sigma N$ coupling constant in
matter. It is a very fundamental issue for QCD whether the nucleon acts
as a dia-scalar. 

On the basis of approaches such as the QMC model, 
there can be little doubt that the
structure of the bound ``nucleon'' is significantly 
different from that of the free nucleon. What occupies the shell-model
orbits in finite nuclei are quasi-particles with nucleon quantum
numbers \cite{finite1}. It will be important to explore further the
corresponding changes in properties (other the the mass) and to check
calculations of such effects against the best available experimental
information.

This is a very exciting time to be working in this field and we eagerly
await the next theoretical and experimental developments.

\section{ACKNOWLEDGEMENTS}

It is a pleasure to thank M. Ericson, P. Guichon, G. Lolos, D. Lu,
K. Maruyama, K. Saito,
K. Tsushima and A. Williams for helpful comments during the preparation
of this report. This work was supported by the Australian
Research Council.

\end{document}